\documentclass[runningheads]{svmult}

\usepackage{makeidx}   
\usepackage{graphicx}  
\usepackage{subeqnar}  
\usepackage{multicol}  
\usepackage{physprbb}  
\makeindex             

\begin{document}
\title*{{\sc mocassin}: 3D photoionisation and dust radiative transfer modelling of PNe}
\toctitle{{\sc mocassin}: 3D photoionisation and dust RT modelling of PNe}
%
%
\titlerunning{{\sc mocassin}}
%
\author{Barbara Ercolano\inst{1}
\and M.~J. Barlow\inst{1}
\and P.~J. Storey\inst{1}
\and X.-W.~Liu\inst{2}
}
\authorrunning{Barbara Ercolano et al.}
%
%
\institute{Department of Physics and Astronomy, University College London, UK
\and Department of Astronomy, Peking University, Beijing, China}
\maketitle              

\begin{abstract}
We present the first 3D Monte Carlo photoionisation code to include a fully self-consistent treatment of dust radiative transfer (RT) within the photoionised region. This is the latest development of the recently published pure photoionisation code {\sc mocassin} (Ercolano et al., 2003a) and it is currently undergoing several benchmarking tests. The preliminary results of these tests are presented in these conference proceedings. The new code provides the ideal tool for the analysis of dusty ionised regions showing asymmetries and/or density and chemical inhomogeneities
\end{abstract}

\section{Introduction}

Recent advances in instrumentation allow the geometry of many PNe (including those in the SMC and LMC) to be resolved. It is now clear that the assumption of spherical symmetry is not justified in most cases. This imposes serious limitations on the application of 1D codes, with the risk of obtaining incorrect abundance determinations, resulting in misleading conclusions. The 3D photoionisation code {\sc mocassin} (MOnte CArlo SimulationS of Ionised Nebulae, Ercolano et al. 2003a) was designed to remedy these serious shortcomings and has been successfully applied to the study of several PNe (Ercolano et al. 2003b,c; Ercolano et al. 2004). \\

The presence of dust grains in ionised environments can have significant effects on the radiative transfer, as the grains compete with the gas for the absorption of UV photons, as well as being heated by nebular resonance line photons. These processes can only be treated properly by incorporating the scattering, absorption and emission of radiation by dust particles that are mixed with the gas in the photoionised region. Moreover, the accurate determination of dust temperatures and spectral energy distributions (SEDs) can only be achieved by treating discrete grain sizes and different species separately.  All of the above has already been implemented in the new version of {\sc mocassin}, which is currently undergoing benchmarking tests, as described below.

\section{Benchmarking Tests}
The dust and gas version of {\sc mocassin} is benchmarked against the popular 1D code, {\sc dusty}, for the spherically symmetric, homogeneous test models described by Ivezic et al. (1997). 
The input parameters for these pure-silicate models are given in their paper and the grain optical constants used are those included in {\sc dusty}'s library, calculated from the data of Draine \& Lee (1984). 


Figure~1 shows the comparison between {\sc mocassin} -- crosses -- and {\sc dusty} (Ivezic et al., 1997) -- dashed line-- for the p=0 (see Ivezic et al., 1997) and $\tau_{1{\mu}m}$~=~10. The emerging SED in relative flux units are shown on the left panel and the radial grain temperature distribution on the right panel. Taking into account that a completely different approach is used by the two codes, the results obtained by the benchmarking are in good agreement. 
2D disk benchmarks are currently being carried out for the pure dust case, whilst a dust and gas model of the PN NGC~3918 is also being constructed (Ercolano et al. in preparation). 

While a number of Monte Carlo dust radiative transfer codes already exist (e.g. Bjorkman \& Wood, 2001), {\sc mocassin} is currently the only 3D code capable of self-consistently treating dust RT within the photoionised region, providing the ideal tool for the modelling of non-spherical, density and/or chemically inhomogeneous dusty photoionised environments. 

\begin{figure*}
\begin{center}
\begin{minipage}[t]{6cm}
\includegraphics[width=6cm]{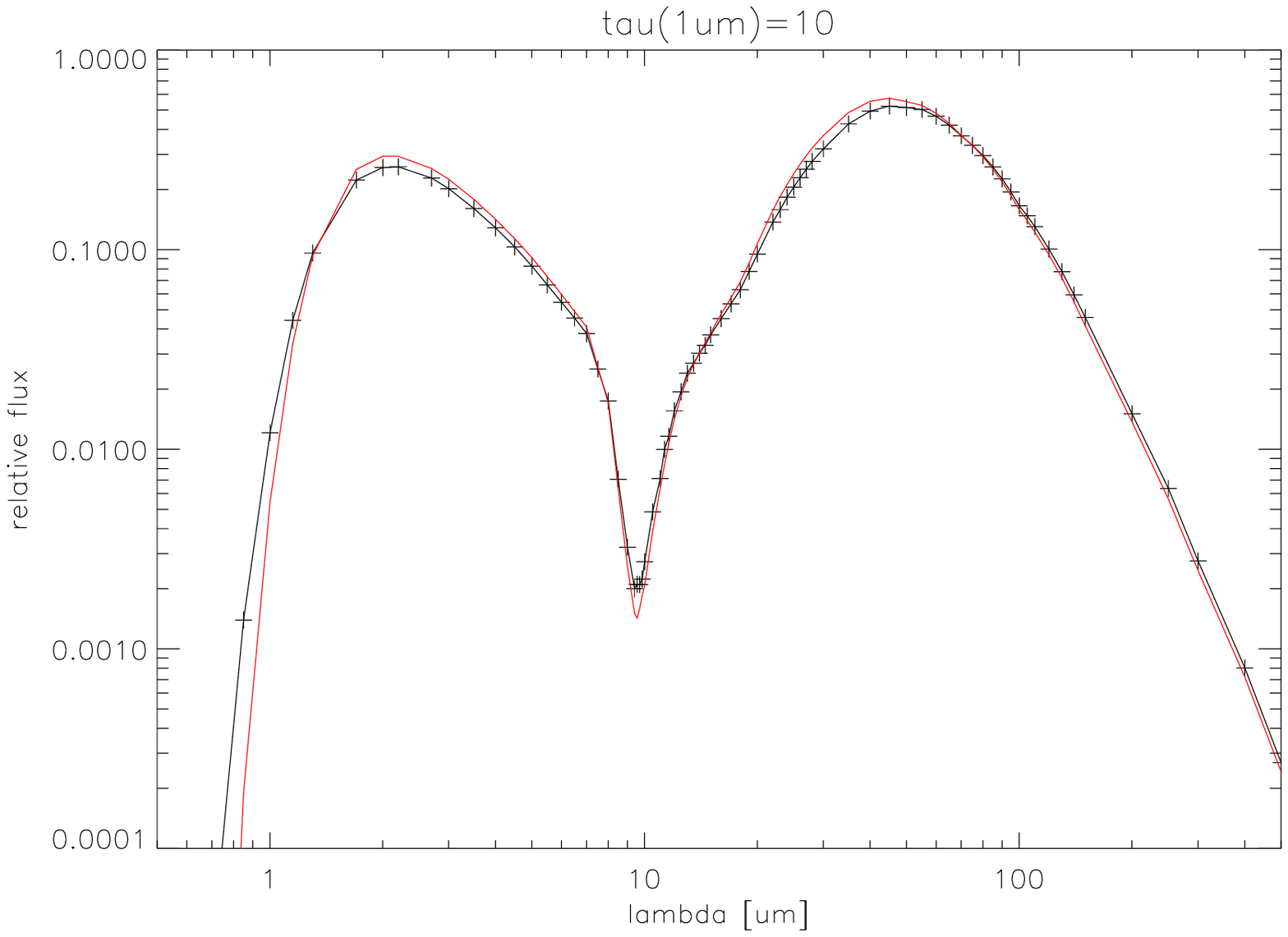}
\end{minipage}
\begin{minipage}[t]{6cm}
\includegraphics[width=6cm]{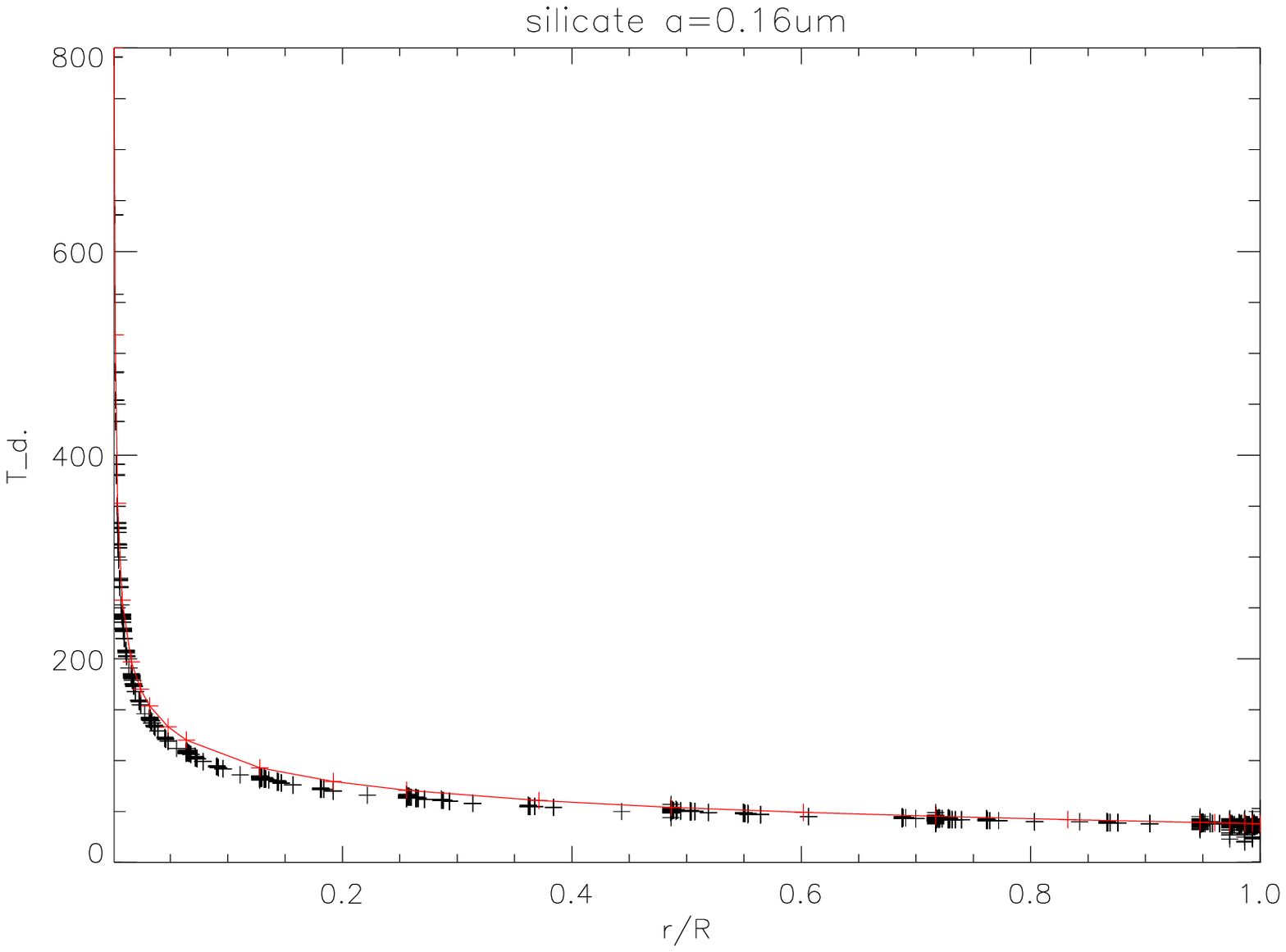}
\end{minipage}
\caption[]{Spherically symmetric, homogeneous benchmark model. See text for details.}
\end{center}
\end{figure*}

%


\begin{thebibliography}{8.}
\addcontentsline{toc}{section}{References}
\bibitem{journ0} J.E. Bjorkman, K. Wood: ApJ 554, 615 (2001)
\bibitem{journ0.5} D.T. Draine, H.M. Lee: ApJ 285, 89 (1984)
\bibitem{journ1} B. Ercolano, M.J. Barlow, P.J. Storey, X.-W. Liu: MNRAS 340, 1136 (2003a)
\bibitem{journ2} B. Ercolano, C. Morisset, M.J. Barlow, P.J. Storey, X.-W. Liu: MNRAS 340, 1153 (2003b)
\bibitem{journ3} B. Ercolano, M.J. Barlow, P.J. Storey, X.-W. Liu, T. Rauch, K. Werner: MNRAS 344, 1145 (2003c)
\bibitem{journ4} B. Ercolano, R. Wesson, Y. Zhang, M.J. Barlow, O. De Marco, T. Rauch, X.-W. Liu: MNRAS submitted (2004)
\bibitem{journ5} Z. Ivezic, M.A.T. Groenewegen, A. Men'shchikov, R., Szczerba: MNRAS 291, 121 (1997)


\end{thebibliography}
\end{document}